\documentclass[aps,prl,preprint,groupedaddress,nofootinbib]{revtex4}

\usepackage{amsmath,latexsym,amssymb}
\usepackage{graphicx}
\usepackage[latin1]{inputenc}

\usepackage{Macros}
\usepackage{nicefrac}
\usepackage{braket}

\usepackage{comment}

\begin{document}

\preprint{hep-th/0511210}
\preprint{CPHT-RR066.1105}

\title{Coset models and D-branes in group manifolds}

\author{Domenico Orlando}
\email[]{domenico.orlando@cpht.polytechnique.fr}

\thanks{Research partially supported by the EEC under the contracts
    \textsc{mext-ct}-2003-509661, \textsc{mrtn-ct}-2004-005104 and
    \textsc{mrtn-ct}-2004-503369.}
\affiliation{ Centre de Physique Théorique de l'École Polytechnique\footnote{Unité mixte du CNRS et de l'École Polytechnique, UMR 7644.}\\
  91128 Palaiseau, France}
\affiliation{Laboratoire de Physique Théorique de l'École Normale
    Supérieure\footnote{Unité mixte du CNRS et de l'École Normale
      Supérieure, UMR 8549.}\\
    24 rue Lhomond, 75231 Paris Cedex 05, France}

\date{\today}

\begin{abstract}
  We conjecture the existence of a duality between heterotic closed
  strings on homogeneous spaces and symmetry-preserving $D$-branes on
  group manifolds, based on the observation about the coincidence of
  the low-energy field description for the two theories. For the
  closed string side we also give an explicit proof of a
  no-renormalization theorem as a consequence of a hidden symmetry and
  infer that the same property should hold true for the higher order
  terms of the \textsc{dbi} action.
\end{abstract}

\maketitle


\newcommand{\mJ}{\ensuremath{\mathcal{J}}}
\newcommand{\mI}{\ensuremath{\mathcal{I}}}
\newcommand{\mK}{\ensuremath{\mathcal{K}}} 

\newcommand{\F}[2]{{f^{#1}_{\phantom{#1}#2}}}

\newcommand{\h}{\ensuremath{\textsc{h}}}
\newcommand{\e}{\ensuremath{\mathrm{e}}}


One of the main technical advantages provided by the study of models
on group manifolds is that the geometrical analysis can be recast in
Lie algebraic terms. At the same time the underlying conformal
symmetry makes it possible to explicitly study the integrability
properties that, in general, allow for extremely nice behaviours under
renormalization. Wess-Zumino-Witten models can be used as starting
points for many interesting models: the main challenge in this case
consists in partially removing the symmetry while retaining as many
algebraic and integrability properties as possible.

In this note we aim at pointing out an analogy (or, as we will say, a
duality) between two -- in principle disconnected -- constructions
based on \textsc{wzw} models: closed string (heterotic) backgrounds
obtained via asymmetric deformations and symmetry-preserving D-branes
on group manifolds.  As we will show, in fact, the low-energy field
contents for both theories are the same, although they minimize
different effective actions (\textsc{sugra} for the former and
\textsc{dbi} for the latter).  For one of the sides of the duality
(the closed string one) we will also show a no-renormalization theorem
stating that the effect of higher-order terms can be resummed to a
shift in the radii of the manifold. A similar behaviour can also be
conjectured from the $D$-brane side, and this would be consistent with
some remark in literature about the coincidence between the
\textsc{dbi} and \textsc{cft} results concerning mass spectra, \ldots
up to the said shift \cite{Bachas:2000ik,Bordalo:2001ec}.

\bigskip 

Let us start with the open-string side of this duality, by reminding
some known facts about the geometric description of
D\nobreakdash-branes in \textsc{wzw} models on compact groups,
pointing out in particular the low-energy field configuration. Natural
boundary conditions on \textsc{wzw} models are those in which the
gluing between left- and right-moving currents can be expressed in
terms of automorphisms $\omega$ of the current algebra. The corresponding
world-volumes are then given by (twisted) conjugacy classes on the
group~\cite{Alekseev:1998mc}:
\begin{equation}
  \mathcal{C}^{\omega} (g) = \Set{h g \omega(h^{-1})| h \in G} .
\end{equation}

As it was pointed out in~\cite{Bachas:2000ik}, one can use Weyl's
theory of conjugacy classes so to give a geometric description of
$C^{\omega}(g)$. For a given automorphism $\omega$ we can always find an
$\omega$-invariant maximal torus $T \subset G $ (such as $\omega(T) = T$).  Let $T^{\omega}
\subset T$ be the set of elements $t \in T$ invariant under $\omega$ ($T^{\omega} =
\Set{t \in T | \omega(t) = t}$) and $T^{\omega}_0 \subset T^{\omega}$ the connected component
to the unity. When $\omega$ is inner $ T = T^{\omega} = T^{\omega}_0$ while in
general (\emph{i.e.} if we allow $\omega$ to be outer) $\dim (T^{\omega}_0) \leq
\rank G$.

Let $\omega$ be inner. Define a map:
\begin{align}
  q: G/T \times T &\to G \nonumber \\
  \left( [g], t \right) &\mapsto q([g],t) = gtg^{-1}.
\end{align}
One can show that this map is surjective, so that each element in $G$ is
conjugated to some element in $T$. This implies in particular that
the conjugacy classes are characterized by elements in $T$, or, in other
words, fixing $t \in T$ (so to take care of the action of the Weyl
group), we find that the (regular) conjugacy classes $\mathcal{C}^{\omega}
(g)$ are isomorphic to the homogeneous space $G/T$. A similar result
holds for twisted classes, but in this case
\begin{equation}
  \mathcal{C}^{\omega} (g) \simeq G/T^{\omega}_0.  
\end{equation}

The description of the D\nobreakdash-brane is completed by the $U(1)$
gauge field that lives on it. The possible $U(1)$ fluxes are elements
in $H^2 ( G / T^\omega_0, \setR)$ and one can show that
\begin{equation}
  H^2 ( G/T^\omega_0 ) \simeq \setZ^{\dim T^\omega_0}.  
\end{equation}

\bigskip

Summarizing we find that the gauge content of the low energy theory is given by:
\begin{itemize}
\item the metric on $G/T^\omega_0$ (in particular $G/T$ for untwisted
  branes),
\item the pull-back of the Kalb--Ramond field on $G/T^\omega_0$,
\item $\dim T^\omega_0$ independent $U(1) $ fluxes ($\rank G$ for
  untwisted branes).
\end{itemize}
These fields extremize the \textsc{dbi} action
\begin{equation}
  \label{eq:dbi}
  S  = \int \di x \: \sqrt{ \det \left( \mathbf{g} + \mathbf{B} + 2 \pi \mathbf{F} \right) }.
\end{equation}
and according to some coincidence with known exact \textsc{cft}
results there are reasons to believe that the fields only receive a
normalization shift when computed at all loops.

\bigskip

Let us now move to the other -- closed string -- side of the
advertised duality.  A good candidate for a deformation of a
\textsc{wzw} model that reduces the symmetry, at the same time
preserving the integrability and renormalization properties, is
obtained via the introduction of a truly marginal operator written as
the product of a holomorphic and an antiholomorphic current
\begin{equation}
  \mathcal{O} = \sum_{ij} c_{ij} J^i \bar J^j .
\end{equation} 
As it was shown in~\cite{Chaudhuri:1989qb}, a necessary and sufficient
condition for this marginal operator to be integrable is that the left
and right currents both belong to abelian groups.  If we consider the
heterotic super\nobreakdash-\textsc{wzw} model, a possible choice
consists in taking the left currents in the Cartan torus and the right
currents from the heterotic gauge
sector~\cite{Kiritsis:1995iu,Israel:2004cd}:
\begin{equation}
  \mathcal{O} = \sum_{a=1}^N \h_a J^a \bar I^a  .
\end{equation}  
where $J^a \in H \subset T$, $T$ being the maximal torus in $G$. 

Using a construction bearing many resemblances to a Kaluza--Klein
reduction it is straightforward to show that the background fields
corresponding to this kind of deformation consist in a metric, a
Kalb--Ramond field and a $U(1)^N$ gauge field. Their explicit
expressions are simply given in terms of Maurer--Cartan one-forms on
$G$ as follows:
\begin{widetext}
  \begin{subequations}
    \label{eq:background-fields}
    \begin{align}
      g &= \frac{k}{2} \delta_{\textsc{mn}} \mJ^{\textsc{m}} \otimes
      \mJ^{\textsc{n}} - k \delta_{ab} \h_a^2 \tilde
      \mJ^a \otimes \tilde \mJ^b  , \label{eq:back-metric}\\
      H_{[3]} &= \di B - \frac{1}{k_g} A^a \land \di A^a = \frac{k}{2}
      f_{\textsc{mnp}}\mJ^{\textsc{m}} \land \mJ^{\textsc{n}} \land
      \mJ^{\textsc{p}} - k \h_a^2 f_{a\textsc{mn}} \mJ^a
      \land \mJ^{\textsc{m}} \land \mJ^{\textsc{n}} ,\\
      A^a &= \h_a \sqrt{\frac{2k}{k_g}} \tilde \mJ^a \text{
        (no summation over $a$ implied)},
    \end{align}
  \end{subequations}
\end{widetext}
where $\tilde \mJ^a_\mu $ are the currents that have been selected for
the deformation operator. In this way we get an $N$-dimensional space
of exact models. Here we will concentrate on a special point in this
space, namely the one that corresponds to $\set{\h_a = 1/ \sqrt{2}, \forall
  a = 1, 2, \ldots, N}$. This point is remarkable for it corresponds to a
decompactification limit where $N$ dimensions decouple and we're left
with the homogeneous $G/H$ space times $N$ non-compact
dimensions\footnote{One can see the initial group manifold $G$ as a
  principal fibration of $H$ over a $G/H$ basis: the deformation
  changes the radii of the fiber and eventually trivializes in
  correspondence of this special point.}.  More precisely, when $H$
coincides with the maximal torus $T$, the background fields read:
\begin{subequations}
\label{eq:coset-background}
  \begin{align}
    G &= \frac{k}{2} \sum_\mu \mJ^\mu \otimes \mJ^\mu , \\
    H_{[3]} &= \di B  = \frac{1}{2} f_{\mu
      \nu \rho} \mJ^\mu \land \mJ^\nu \land \mJ^\rho ,\\
    F^a &= - \sqrt{\frac{k}{2k_g}} \h_a \F{a}{\mu \nu} \mJ^\mu \land
    \mJ^\nu
  \end{align}
\end{subequations}
(no summation over $a$).
Geometrically:
\begin{itemize}
\item $g$ is the metric on $G/T$ obtained as the restriction of the
  Cartan--Killing metric on $G$
\item $H_{[3]}$ is the pullback of the usual Kalb--Ramond field present in
  the \textsc{wzw} model on the group $G$
\item $F^a$ are $\rank(G)$ independent $U(1)$ gauge fluxes that
  satisfy some quantization conditions and hence naturally live in
  $H^2 ( G/T, \setZ ) $
\end{itemize}

Having chosen a truly marginal operator for the deformation we know
that this model is conformal. This implies in particular that the
background fields solve the usual $\beta $ equations that stem from
the variation of the effective \textsc{sugra} action:
\begin{equation}
  \label{eq:sugra}
  S = \int \di x \: \sqrt{g} \left( R - \frac{1}{12} H_{\mu \rho \sigma} 
    H^{\mu \rho \sigma} - \frac{k_g}{8} F^a_{\mu \nu} F^{a\mu \nu } 
    + \frac{\delta c}{3} \right)
\end{equation}

In example if we consider $G= SU (2)$, then $T = U(1)$ and the
decompactification limit $H \to 1/\sqrt{2}$ we get the exact $S^2 = SU
(2)/U(1)$ background supported by a $U(1)$ magnetic monopole field
(see \emph{e.g.}~\cite{Johnson:1995kv,Berglund:1996dv}).

Our conjecture stems precisely from this: the gauge field above
exactly match the ones we found before for symmetry-preserving
D-branes. Moreover both sides of the duality are derived from
\textsc{wzw} models that enjoy a no-renormalization property which
would make this correspondence true at all orders. In this spirit we
now pass to prove that a similar theorem holds for closed heterotic
strings on coset models infering that the duality, when proven, would
give a direct way to deduce the same feature  for the D-brane action.

\bigskip
%

In studying symmetrically deformed \textsc{wzw} models, \emph{i.e.}
those where the deformation operator is written as the product of two
currents belonging to the same sector $\mathcal{O} = \lambda J \bar
J$, one finds that the Lagrangian formulation only corresponds to a
small-deformation approximation. For this reason different techniques
have been developed so to read the background fields at every order in
$\lambda
$~\cite{Hassan:1992gi,Giveon:1994ph,Forste:1994wp,Forste:2003km,Detournay:2005fz}
but, still, the results are in general only valid at first order in
$\alpha^\prime $ and have to be modified so to take into account the
effect of instanton corrections. In this section we want to show that
this is not the case for asymmetrically deformed models, for which the
background fields in Eqs.~\eqref{eq:background-fields} are exact at
all orders in $\h_a$ and for which the effect of renormalization only
amounts to the usual (for \textsc{wzw} models) shift in the level of
the algebra $k \to k + c_G$ where $c_G$ is the dual Coxeter number.

Consider in example the most simple $SU(2)$ case. In terms of Euler
angles the deformed Lagrangian is written as:
\begin{widetext}
  \begin{multline}
    S = S_{SU(2)} \left( \alpha, \beta, \gamma \right) + \delta S = \frac{k}{4 \pi } \int
    \di^2 z \: \d \alpha \db \alpha + \d \beta \db \beta + \d \gamma \db \gamma + 2 \cos \beta \d \alpha
    \db \gamma + \\+ \frac{\sqrt{k k_g} \h}{2 \pi } \int \di^2 z \: \left( \d \gamma +
      \cos \beta \d \alpha \right) \bar I .
  \end{multline}
  If we bosonize the right-moving current as $\bar I = \bar \partial \phi$ and add a
  standard $U(1) $ term to the action, we get:
  \begin{multline}
    S = S_{SU(2)} \left( \alpha, \beta, \gamma \right) + \delta S \left( \alpha, \beta, \gamma, \phi
    \right) + \frac{k_g}{4 \pi } \int \di^2 z \: \d \phi \db \phi =\\= S_{SU(2)} \left(
      \alpha, \beta, \gamma + 2 \sqrt{\frac{k_g}{k}} \h \phi \right) + \frac{k_g \left( 1
        -2 \h^2\right)}{4 \pi} \int \di^2 z \d \phi \db \phi
  \end{multline}
\end{widetext}
and in particular at the decoupling limit $\h \to 1/\sqrt{2} $,
corresponding to the $S^2$ geometry, the action is just given by $ S=
S_{SU(2)} \left( \alpha, \beta, \gamma + 2 \sqrt{\frac{k_g}{k}} \h \phi \right)$. This
implies that our (deformed) model inherits all the integrability and
renormalization properties of the standard $SU (2) $ \textsc{wzw}
model. In other words the three-dimensional model with metric and
Kalb--Ramond field with $SU(2) \times U(1)$ symmetry and a $U(1)$ gauge
field is uplifted to an exact model on the $SU(2)$ group manifold (at
least locally): the integrability properties are then a consequence of
this hidden $SU(2)\times SU(2)$ symmetry that is manifest in higher
dimensions.

The generalization of this particular construction to higher groups is
easily obtained if one remarks that the Euler parametrization for the $g \in SU
\left( 2 \right)$ group representative is written as:
\begin{equation}
  g = \e^{\imath \gamma t_3} \e^{\imath \beta t_1} \e^{\imath \alpha t_2} , 
\end{equation}
where $t_i = \sigma_i /2 $ are the generators of $\mathfrak{su}( 2)$ ($\sigma_i$ being
the usual Pauli matrices). As stated above, the limit deformation
corresponds to the gauging of the left action of an abelian subgroup $T \subset
SU \left( 2\right)$. In particular here we chose $T = \set{h | h = \e^{\imath \phi
  t_3}}$, hence it is natural to find (up to the normalization) that:
\begin{equation}
  h \left( \phi \right) g \left( \alpha, \beta, \gamma \right) = g \left( \alpha, \beta, \gamma + \phi
  \right) .
\end{equation}
The only thing that one needs to do in order to generalize this result
to a general group $G$ consists in finding a parametrization of $g \in
G$ such as the chosen abelian subgroup appears as a left factor. In
example if in $SU(3) $ we want to gauge the $U \left( 1 \right)^2$
abelian subgroup generated by $\braket{\lambda_3, \lambda_8}$
(Gell-Mann matrices), we can choose the following parametrization for
$g \in SU (3)$~\cite{Byrd:1997uq}:
\begin{equation}
  g = \e^{\imath \lambda_8 \phi } \e^{\imath \lambda_3 c } \e^{\imath \lambda_2 b } \e^{\imath \lambda_3 a }
  \e^{\imath \lambda_5 \vartheta } \e^{\imath \lambda_3 \gamma  } \e^{\imath \lambda_2 \beta  } \e^{\imath \lambda_3 \alpha  } . 
\end{equation}

The deep reason that lies behind this property (differentiating
symmetric and asymmetric deformations) is the fact that not only the
currents used for the deformation are preserved (as it happens in both
cases), but here their very expression is just modified by a constant
factor. In fact, if we write the deformed metric as in
Eq.~\eqref{eq:back-metric} and call $\tilde K^\mu $ the Killing vector
corresponding to the chosen isometry (that doesn't change along the
deformation), we see that the corresponding $\tilde \mJ_\mu^{(\h)} $
current is given by:
\begin{equation}
  \tilde \mJ_\nu^{(\h)} = \tilde K^\mu g_{\mu \nu }^{(\h)} = \left( 1 -
    2 \h^2\right) \tilde \mJ_\nu^{(0)}
\end{equation}
The most important consequence (from our point of view) of this
integrability property is that the \textsc{sugra} action in
Eq.~(\ref{eq:sugra}) is \emph{exact} and the only effect of
renormalization is the $k \to k + c_G $ shift.

It is very tempting to exend this no-renormalization theorem to the
$D$-brane side. Of course this would require an actual proof of the
duality we conjecture. Nevertheless we think that this kind of
approach might prove (at least for these highly symmetric systems)
more fruitful than adding higher loop corrections to the \textsc{dbi}
action, which on the other hand remains an interesting directions of
study by itself.

  \bigskip



Is this duality just a coincidence, due to the underlying Lie
algebraic structures that both sides share, or is it a sign of the
presence of some deeper connection? Different aspects of the profound
meaning of the \textsc{dbi} effective action are still poorly
understood and it is possible that this approach -- pointing to one
more link to conformal field theory -- might help shedding some new
light.

\bigskip


\bigskip

\begin{acknowledgments}
  I would like to thank C.~Bachas, C.~Kounnas and S.~Ribault for
  illuminating discussions and especially thank M.~Petropoulos for
  encouragement during different stages of this work and for the
  careful reading of the manuscript.
\end{acknowledgments}

\bibliography{Biblia}

\end{document}